\documentstyle[12pt]{article}
\def\noin{\noindent}
\def\bib{\bibitem}

\def\bea {\begin{eqnarray}}
\def\eea {\end{eqnarray}}
\def\beas{\begin{eqnarray*}}
\def\eeas{\end{eqnarray*}}
\def\be{\begin{equation}}
\def\ee{\end{equation}}

\def\PRL{Phys. Rev. Lett.}

\def\be{\begin{equation}}
\def\ee{\end{equation}}
\def\bea{\begin{eqnarray}}
\def\eea{\end{eqnarray}}

\def\noin{\noindent}

\def\lab{\label}

\def\eg{{\it e.g.\ }}

\def\lsim{\mbox{{\scriptsize \raisebox{-.9ex}
      {$\;\stackrel{{\textstyle <}}{\sim}\,$} }} }

\newcommand{\e}{{\mbox{e}}}
\def\CPT{{\small $\chi$PT}}

\def\mN{m_{\mbox{\tiny N}}}

\def\lsim{\mbox{{\scriptsize \raisebox{-.9ex}
      {$\;\stackrel{{\textstyle <}}{\sim}\,$} }} }

\renewcommand{\thefootnote}{\fnsymbol{footnote}}

\begin{document}


\begin{center}
{\bf  Effective Field Theory and High-Precision Calculations \\ of
Nuclear Electroweak Processes}\footnote{To appear in {\it Festschrift for Gerry Brown}, ed. Sabine Lee (World Scientific, Singapore)}

\vspace{1cm}

Kuniharu Kubodera$^1$ and Mannque Rho$^2$

\vspace{0.5cm}
$^1$ Department of Physics and Astronomy,\\
University of South Carolina,\\
Columbia, SC 29208, USA

\vspace{3mm}

$^2$ Institut de Physique Th\'{e}orique, CEA Saclay,\\
91191 GIf-sur-Yvette C\'{e}dex, France, \&\\
Department of Physics, Hanyang University,\\
Seoul 133-791, Korea

\end{center}

\vspace{1cm}

\centerline{\bf Abstract}

\vskip 0.5cm
\noin
High-precision calculations
of electroweak processes in light nuclei
are of great importance for multiple reasons.
Gerry Brown together with Dan-Olof Riska
published in 1972 a famous calculation
on radiative capture of a thermal neutron on a proton,
and their work was a precursor
of subsequent great developments
in high-precision calculations of
electroweak processes in light nuclei.
The application of effective field theory
to nuclear systems is a prominent example of
these developments.
We present an overview of it
with main emphasis placed
on those topics in which we ourselves have been involved.

\setcounter{footnote}{0}
\renewcommand{\thefootnote}{\arabic{footnote}}
\newpage

\section{Introduction}
In 1972 Dan-Olof Riska and Gerry Brown published
their celebrated calculation~\cite{RB72}
on radiative capture of a thermal neutron on a proton:
$np\to\gamma d$.
Until their work appeared,
the $\sim$10 \% discrepancy
between the experimental and theoretical values
of the radiative $np$ capture rate
had been a persistent puzzle.
Dan-Olof and Gerry demonstrated
that this discrepancy could be removed
by including the contributions of the exchange currents
derived by Chemtob and Rho~\cite{CR71},\footnote{see M. Rho,
in this volume.} and this was the first convincing case
that pointed to the existence of the exchange currents
in nuclei. The 1972 Riska-Brown paper was a precursor
of subsequent important developments
in high-precision calculations of
electroweak processes in light nuclei.
On the occasion of Gerry's 85th birthday,
we would like to give a succinct survey of
these developments;
we shall concentrate here mostly on those topics
we ourselves have worked on,
and so this note is not meant to be a
comprehensive survey of the field.

High-precision calculations
of electroweak processes in light nuclei
are of great importance for multiple reasons.
First of all, light nuclei, for which nuclear many-body-problem aspects are
relatively well controlled, are expected to provide clean testing grounds
for our understanding of the behavior of the hadronic electroweak currents
in nuclei.
In this connection, we should emphasize
that the recent progress in formalisms and numerical techniques
for solving the multi-nucleon Schr\"{o}dinger equation
has reached such a level \cite{cs98} that the wave functions
of low-lying levels for light nuclei
can now be obtained practically with no approximation
(once the validity of a given nuclear Hamiltonian is assumed).
This situation frees us from the
``usual" nuclear physics complications
that arise as a result of
truncating the nuclear Hilbert space
down to certain model space
(such as shell-model configurations,
cluster-model trial functions, RPA, etc.).

We should also mention that nuclear electroweak processes
in light nuclei are of particular importance in connection
with astrophysics and particle physics.
They appear in many astrophysical phenomena,
prominent examples being stellar evolution
(including solar nuclear fusion), big-bang nucleo-synthesis,
supernova nucleo-synthesis, etc.
Furthermore, in terrestrial experiments designed to detect
astrophysical neutrinos, weak-interaction reactions on light nuclei
play important roles.
It is also to be noted that neutrino-nucleus reactions feature
importantly in neutrino-oscillation experiments,
one of the most hotly pursued frontiers in particle physics.
Reliable information on the cross sections for these processes
is obviously in great demand.
However, for many of the processes in question,
it is extremely difficult, or even impossible at present,
to experimentally measure their cross sections.
We must therefore totally rely on theoretical estimates,
and this implies that we need a theoretical framework that allows
us to arrive at estimates with reliable error assessments attached.
Fortunately, the past two decades have witnessed significant progress
toward this goal.

Recent developments in high-precision calculations
of electroweak reactions in light nuclei
encompass two lines of studies.
One is the elaboration and expansion of
the phenomenological potential model,
which has come to be known
as the standard nuclear physics approach (SNPA).
Another line is the successful application of effective field theory (EFT)
to nuclear physics.
In the sense that the former is phenomenological in nature
whereas the latter is based on a systematic expansion
that is formally in conformity with the tenets of EFT,
these two approaches are distinct
and, perhaps because of this,  it happens rather often
that practitioners of one approach (SNPA or EFT)
ignore the existence of and progress in the other approach.
In our view, however, this does not seem an ideal or even productive situation.
As discussed below, the forte of EFT is that  it
gives a clear-cut guiding principle on how to tame short-range physics.
But, as far as long-range nuclear correlations are concerned,
EFT does not necessarily lead to improvement over what has already been
achieved in SNPA.
And, as we will illustrate with concrete examples later,
the high-precision calculation of an observable
often requires that both short-range physics and long-range nuclear correlations
are under control.
So, at least at present, it appears to be a reasonable strategy
to take advantage of the availability
and the power
of both SNPA and EFT
while assuring,  as best one can, consistency between the two.
This viewpoint led us to propose
the ``more effective"  effective field theory (MEEFT) approach,
also called EFT*.
It should be emphasized here that Tae-Sun Park
has played a particularly important role in the developments of this approach.
The nature of EFT limits us to consider
only those processes which involve relatively low energies and momenta
but, within this limitation,
MEEFT has registered remarkable successes
in giving reliable estimates for electroweak transition rates
of current interest.
We wish to describe here some of these results;
primarily we shall be concerned with the following processes:

$pp$-fusion:
\hspace{1.6cm}
$p+p\to d +\!e^+\!\!+\! {\nu_e}$

$\nu$-$d$ reactions:
\hspace{0.8cm}
$\nu_e\!+d \to e^-\!\!+p\!+\!p$,\,\,\,
$\nu+d \to \nu\!+\!n\!+\!p$

$\mu$-$d$ capture:
\hspace{1cm}
$\mu^-\!+\! d \to \nu_\mu\!+\!n\!+\!n$

Radiatice $\pi$-capture:
\hspace{2mm} $\pi^- \!+d \to\, \gamma+ n+n$.

Tritium $\beta$-decay:
\hspace{0.6cm}
$t\to \,^3{\rm He}+e^-\!+\bar{\nu}_e$

$\mu$ -$^3{\rm He}$ capture:
\hspace{1cm}
$\mu^-\!+ ^3\!{\rm He} \to \nu_\mu+t$

Hep reaction
\hspace{1.3cm}$^3{\rm He}+\!p
\to \,^4{\rm He}+\!\nu_e\!+\! e^+ $

Hen reaction
\hspace{1.3cm}$^3{\rm He}+\!n \to \,^4{\rm He}+\!\gamma $

\vspace{2mm}
\noin
This list may look like a haphazardly picked collection
but, as will be explained below,
they are interrelated in the context of demonstrating
the important progress achieved in MEEFT.
Specific motivations for considering these reactions
will be described later for each case as
the individual processes are discussed.
Just to set the stage, however,
we make some remarks here on the following three processes:
$pp$-fusion, Hep and the $\nu d$ reaction.

The  $pp$ fusion reaction, $pp\to d\,e^+{\nu_e}$,
is the primary solar nuclear reaction
that essentially controls the burning rate of the sun,
and hence the exact value of its cross section is a crucial
input for any further developments of solar models.
According to the latest review~\cite{ADEetal2010},
this rate is required to be known with better than $\sim$1 \% precision
in order to give impact to the existing models.
Meanwhile, the Hep reaction,
$^3{\rm He}\!+\!p \rightarrow ^4\!{\rm He}\!+\!e^+\!+\!\nu_e$
is important in a different context.
Although Hep plays only a very minor role
in the pp-chain solar burning,
it generates highest-energy solar neutrinos
whose spectrum extends beyond the maximum energy
of the $^8$B neutrinos.
So, even though the flux of the Hep neutrinos
is small, it can distort the higher end of
the $^8$B neutrino spectrum~\cite{kuz65}, and this distortion
can affect the interpretation of the
results of a Super-Kamiokande experiment~\cite{SuperK-Hep}.
On the theoretical side,
as explained in more detail below,
Hep represents a unique case in which
the leading-order contribution is drastically suppressed.
This implies that Hep offers a good place for investigating
higher order terms in EFT expansion.
As for the $\nu$-$d$ reactions, their primary importance
is connected to the Sudbury Neutrino Observatory (SNO) experiments,
in which the solar neutrino flux was monitored
using a  heavy water Cerenkov counter.
As is well known, the SNO experiments~\cite{ahmetal}
have established that
the total solar neutrino flux
(summed over all neutrino flavors) agrees with the prediction
of the standard solar model~\cite{bahcall},
whereas the electron neutrino flux
from the sun is significantly smaller
than the total solar neutrino flux.
These results of the SNO experiments
have given ``smoking-gun" evidence
for the transmutation of solar electron neutrinos into
neutrinos of other flavors.
Obviously, a precise knowledge
of the $\nu$-$d$ reaction cross sections
is of primary importance
in interpreting the SNO data.

Before going into the description of achievements
obtained in nuclear EFT,
it seems useful to take a quick look
at the latest lattice QCD calculation~\cite{YAMetal09}
of the nucleon weak form factors defined as
\bea
<\!n|V^\mu|p\!> &=&
\bar{u}_n\!\left[F_V(q^2)\gamma^\mu
\!+\frac{iF_M(q^2)}{2m_N}\sigma^{\mu\nu}q_\nu
\right]
u_p\,,\nonumber \\
<\!n|A^\mu|p\!> &=&
\bar{u}_n\left[F_A(q^2)\gamma^\mu\gamma^5
\!+\frac{F_P(q^2)}{m_\mu}q^\mu\gamma^5
\right]
\!u_p\,, \label{eq:Afi}
\eea
where $q=p_n-p_p$.
The calculation~\cite{YAMetal09}
was carried out for the $u/d$ quark mass
that corresponded to $m_\pi$ = 0.33 GeV $\sim$ 0.67 GeV,
and the form factors were calculated for $0.2$ GeV$^2<q^2<0.75$ GeV$^2$.
Here is a partial summary of the results:
(1)  $F_V(q^2)$ is described by the conventional dipole form,
but the dipole mass $M_1$
is found to be 27~\% smaller than the empirical value;
(2) $F_M(q^2)$ is described by the dipole form,
but the corresponding dipole mass $M_2$
is found to be 26~\% smaller than the empirical value;
(3) $g_A/g_V=1.19$,
which is 7~\% smaller than the experimental value
$(g_A/g_V)_{\rm exp} = 1.2695\pm0.0029$ (PDG08);
(4) $F_A(q^2)$ is found to have the dipole form
but the axial dipole mass $M_A$ is found to be 37~\% larger
than $M_A^{\rm exp}$;
(5) $\frac{(q^2-m_\pi^2)F_P(q^2)}{2m_NF_A(q^2)}
\equiv \alpha(q^2)$
is found to be constant in conformity with pion-pole dominance,
but $\alpha=0.8\sim 0.9$ (instead of $\alpha=1$ expected from PCAC).
These features lead us to expect that
it
will be a while before nuclear elecroweak processes
can be calculated with required precision by lattice QCD.

\section{Theoretical frameworks}

\noin
$\bullet$ \hspace{1mm}
{\bf Standard nuclear physics approach (SNPA)}

The phenomenological potential
picture has been highly successful in describing
many different nuclear phenomena.
In this picture an $A$-nucleon system
is described by a Hamiltonian of the form
\be
H\,=\,\sum_i^A t_i + \sum_{i<j}^A V_{ij}
+\sum_{i<j<k}^A V_{ijk}+ \cdots\,, \label{HSNPA}
\ee
where $t_i$ is the kinetic energy of the $i$-th nucleon,
$V_{ij}$ is a phenomenological two-body potential,
$V_{ijk}$ is a phenomenological three-body potential,
and so on.
Once $H$ is specified,
the $A$-body nuclear wave function $|\Psi\!>$
is obtained by solving {\it exactly or essentially exactly}
the Schr\"{o}dinger equation
\be
H|\Psi\!>\,=\,E|\Psi\!>\,.\label{Sch}
\ee
There is much freedom in choosing
possible forms of $V_{ij}$
apart from a well-established requirement that,
as the inter-nucleon distance
$r_{ij}$ becomes sufficiently large,
$V_{ij}$ should approach the one-pion exchange
Yukawa potential.
For the model-dependent short-range part
of $V_{ij}$, one assumes certain functional forms
and fix the parameters appearing therein
by demanding that the solutions of eq.(\ref{Sch})
for the $A$=2 case reproduce the nucleon-nucleon
scattering data
(typically up to the pion-production threshold energy)
as well as the deuteron properties.
There exist several so-called {\it  high-precision}
{\it phenomenological} N-N potentials
that can reproduce all the existing two-nucleon data
with normalized $\chi^2$ values close to 1.
These potentials differ significantly
in the ways they parameterize
short-range physics
and, as a consequence,
they exhibit substantial difference
in their off-shell behavior.
An important question therefore is to what extent
the model-dependent short-range part affects
the reliabilities of theoretical values of various observables
in $A$=2 and higher systems calculated with these phenomenological
potentials.

Nuclear responses to external electroweak probes
in normal circumstances are given, to good approximation,
by one-body terms, which are also called
the impulse approximation (IA) terms.
To achieve higher accuracy, however,
we must also consider exchange current (EXC) terms,
which represent the contributions
of nuclear responses involving two or more nucleons.
If for some reason the IA contributions are suppressed,
it becomes particularly important to consider the EXC contributions.
These exchange currents are usually taken to be two-body operators,
and they are derived from one-boson exchange diagrams;
the vertices featuring in the relevant diagrams
are determined to satisfy the requirements of low-energy theorems
and current algebra \cite{CR71,IT79}.
A formalism based on this picture is commonly called
the {\it standard nuclear physics approach} (SNPA).
(This is also called a potential model in the literature.)
Schematically, an electroweak nuclear matrix element in SNPA
is given by
\be
{\cal M}_{fi}^{SNPA}\,=\,
<\!\Psi_{\!f}\,|\sum_\ell^A{\cal O}_\ell
+\sum_{\ell<m}^A{\cal O}_{\ell m}
+\sum_{\ell<m<n}^A\!\!{\cal O}_{\ell mn} \,
\,|\Psi_i\!>\,, \label{ME-SNPA}
\ee
where the initial (final) nuclear wave function,
$\Psi_i$ ($\Psi_f$), is a solution of eq.(\ref{Sch}),
and ${\cal O}_\ell$, ${\cal O}_{\ell m}$
and ${\cal O}_{\ell mn} $ are, respectively,
one-body,  two-body and three-body transition operators for
a given electroweak  process.
SNPA has been used extensively
for describing electroweak processes
in light nuclei,
and general good agreement found
between theory and experiment~\cite{cs98}
strongly suggests that SNPA captures much
of the physics involved.

\vspace{4mm}
\noin
$\bullet$ \hspace{3mm} {\bf Effective field theory (EFT)}

Even though SNPA has been scoring great successes
in describing a vast variety of data,
it is still important
from the fundamental point of view
to raise the following issues.
First, given the fact that hadrons and hadronic systems
(such as nuclei) are governed
by quantum chromodynamics (QCD),
it is desirable to establish connections
between SNPA with QCD.
In this connection we note that,
whereas chiral symmetry is known
to be a fundamental symmetry of QCD,
the formulation of SNPA is largely disjoint from
this symmetry.
Secondly, in introducing a phenomenological SNPA Lagrangian
for describing the nuclear interaction and
nuclear responses to the electroweak currents,
it is not clear whether there
is any identifiable expansion parameter
that helps us to control the possible forms of terms in
the Lagrangian and that gives a general
measure of errors in our calculation.

An EFT-based approach~\cite{wei90}
is a powerful tool to address these issues,
and it has been studied with great intensity;
for reviews, see~\cite{bkm}-\cite{br02}.
The basic idea of EFT is simple.
To describe phenomena
characterized by a typical energy-momentum scale $Q$,
we expect that we can exclude from our Lagrangian
those degrees of freedom
that belong to energy-momentum scales
much higher than $Q$.
So we introduce a cut-off scale $\Lambda$
that is sufficiently larger than $Q$
and we categorize our fields
(to be generically represented by $\phi$)
into two groups: high-frequency fields
$\phi_{\mbox{\tiny H}}$
and low-frequency fields $\phi_{\mbox{\tiny L}}$.
By {\it integrating out} $\phi_{\mbox{\tiny H}}$,
we arrive at an {\it effective} Lagrangian
that only contains $\phi_{\mbox{\tiny L}}$
as explicit dynamical variables.
The effective Lagrangian ${\cal L}_{{\rm eff}}$
is related to the original Lagrangian ${\cal L}$ as
\be
\int\![d\phi]\e^{{\rm i}\int d^4x{\cal L}[\phi]}
= \int\![d\phi_{\mbox{\tiny H}}][d\phi_{\mbox{\tiny L}}]
\e^{{\rm i}\int d^4x{\cal L}
[\phi_{\mbox{\tiny H}},\phi_{\mbox{\tiny L}}]}
\equiv
\int\![d\phi_{\mbox{\tiny L}}]
\e^{{\rm i}\int d^4x{\cal L}_{\rm eff}
 [\phi_{\mbox{\tiny L}}]}\,.\label{EFTdef}
\ee
${\cal L}_{{\rm eff}}$ defined
by eq.(\ref{EFTdef}) inherits the symmetries
(and the pattern of symmetry breaking, if there is any)
of the original Lagrangian ${\cal L}$.
It can be shown that
${\cal L}_{{\rm eff}}$ is given by
the sum of all possible monomials of
$\phi_{\mbox{\tiny L}}$ and their derivatives
that are consistent with the symmetry requirements
dictated by ${\cal L}$.
Because a term involving $n$ derivatives
scales like $(Q/\Lambda)^n$,
the terms in ${\cal L}_{{\rm eff}}$ can be organized
into a perturbative series
wherein $Q/\Lambda$ serves as an expansion parameter.
The coefficients of terms in
this expansion are called
low-energy constants (LECs).
If all the LEC's up to a specified order $n$
can be fixed either from theory or from fitting
to the experimental values of relevant observables,
${\cal L}_{{\rm eff}}$ serves as
a complete (and hence model-independent) Lagrangian
to the given order of expansion.
The results obtained are ensured to
have accuracy of order $(Q/\Lambda)^{n+1}$.
In applying EFT to nuclear systems,
the underlying Lagrangian ${\cal L}$
is the QCD Lagrangian ${\cal L}_{QCD}$,
whereas, for the typical nuclear physics
energy-momentum scale
$Q\ll \Lambda_{\chi}\sim 1$ GeV,
the effective degrees of freedom
that should feature in ${\cal L}_{{\rm eff}}$
are hadrons rather than the quarks and gluons.
It is not trivial
to apply the formal definition in eq.(\ref{EFTdef})
to derive ${\cal L}_{{\rm eff}}$
written in terms of hadrons
starting from ${\cal L}_{QCD}$;
this is because the hadrons cannot be simply
identified with the low-frequency field,
$\phi_L$ in eq.(\ref{EFTdef}).
At present, the best one could do is
to resort to symmetry considerations
and the above-mentioned expansion scheme
in the spirit of
Weinberg's folk theorem~\cite{wei79}.
Here chiral symmetry plays an important role.
We know that chiral symmetry is spontaneously
broken, generating the pions
as Nambu-Goldstone bosons.\footnote{
We limit ourselves here to SU(2)$\times$SU(2)
chiral symmetry.}
This feature can be taken into account
by assigning suitable chiral transformation properties
to the Goldstone bosons
and writing down all possible chiral-invariant
terms up to a specified chiral order \cite{geo84}.
The above consideration presupposes
exact chiral symmetry in ${\cal L}_{QCD}$.
In reality, however, ${\cal L}_{QCD}$ contains
small but finite quark mass terms,
which explicitly break chiral symmetry
and lead to a non-vanishing value
of the pion mass $m_\pi$.
Again, there is a well-defined method
to determine what terms are needed
in the Goldstone boson sector
to represent the effect of
explicit chiral symmetry breaking
\cite{geo84}.
These considerations lead to an EFT
called chiral perturbation theory
($\chi$PT)~\cite{wei79,gl84}.
The successes of $\chi$PT in
the meson sector are well documented;
see, {\it e.g.,} \cite{bkm}.

A serious problem we encounter in extending $\chi$PT
to the nucleon sector is that,
as the nucleon mass $\mN$ is comparable to
the cut-off scale $\Lambda_{\chi}$,
a simple application of expansion in $Q/\Lambda$
does not work.
This difficulty can be circumvented by employing
heavy-baryon chiral perturbation theory (HB$\chi$PT),
which essentially consists in shifting
the reference point of the nucleon energy
from 0 to $\mN$ and in integrating out
the small component of the nucleon field
as well as the anti-nucleonic degrees of freedom.
Thus an effective Lagrangian in HB$\chi$PT contains
as explicit degrees of freedom
the pions and the large components
of the redefined nucleon field.
HB$\chi$PT has as expansion parameters
$Q/\Lambda_{\chi}$, $m_\pi/\Lambda_{\chi}$
and $Q/\mN$.
As $\mN\approx \Lambda_{\chi}$,
it is convenient to combine chiral
and heavy-baryon expansions
and introduce the chiral index ${\bar \nu}$
defined by $\bar{\nu}=d+(n/2)-2$.
Here $n$ is the number of fermion lines
that participate in a given vertex,
and $d$ is the number of derivatives
(with $m_\pi$ counted as one derivative).
A similar power counting scheme can
be introduced for Feynman diagrams as well~\cite{wei90}.
The contribution of
a Feynman diagram that contains $N_A$ nucleons,
$N_E$ external fields,
$L$ loops and $N_C$ disjoint parts
is shown to scale like
$(Q/\Lambda)^\nu$, where
the chiral index $\nu$ is defined as
\be
\nu = 2 L + 2 (N_C-1) + 2 - (N_A+N_E) +
\sum_i \bar \nu_i\,,\lab{eq:nu}
\ee
with the summation running over all the vertices
contained in the Feynman diagram.
HB$\chi$PT has been applied with great success
to the one-nucleon sector \cite{bkm}.

However, HB$\chi$PT cannot be applied
in a straightforward manner
to nuclei, which contain more than one nucleon.
The reason is that nuclei involve
very low-lying excited states,
and the presence of this small energy scale
spoils the original counting rule~\cite{wei90}.
A method to avoid this difficulty
was proposed by Weinberg~\cite{wei90}.
Classify Feynman diagrams into two categories,
irreducible and reducible diagrams.
Irreducible diagrams are those
in which every intermediate state
has at least one meson in flight;
all others are classified as reducible diagrams.
One then applies the above chiral counting rules
only to irreducible diagrams.
The contributions of all the two-body irreducible diagrams
(up to a specified chiral order)
are treated as an effective potential
(to be denoted by $V_{ij}^{EFT}$)
acting on nuclear wave functions.
Meanwhile, the contributions of reducible diagrams
can be incorporated~\cite{wei90}
by solving the Schr\"odinger equation
\be
H^{EFT}|\Psi\!>^{EFT}\,=\,E|\Psi\!>^{EFT}\,,
\label{Sch-EFT}
\ee
where
\be
H^{EFT}\,=\,\sum_i^A t_i \,+\, \sum_{i<j}^A V_{ij}^{EFT}
+\!\sum_{i<j<k}^A V_{ijk}^{EFT}+\ldots\,,
\label{HEFT}
\ee
We may refer to this two-step procedure as
{\it nuclear} \CPT\ or, more specifically,
{\it nuclear} \CPT\
in the Weinberg scheme.\footnote{
This is often called the $\Lambda$-counting scheme
\cite{lep99}.}

To apply nuclear \CPT\ to a nuclear electroweak process,
we derive a nuclear transition operator
${\cal T}$
from a complete set of
all the irreducible diagrams
(up to a given chiral order $\nu$) involving
the relevant external current(s)~\cite{Rho91,pmr93,pmr95}.
Thus
\be
{\cal T}^{\mbox{\tiny{EFT}}}
= \sum_i^A{\cal O}_i^{\mbox{\tiny{EFT}}}
+\sum_{i<j}^{A}{\cal O}_{ij}^{\mbox{\tiny{EFT}}} +\, \cdots
\ee
To preserve consistency in chiral counting,
the  matrix element of ${\cal T}^{\mbox{\tiny{EFT}}}$
must be calculated with the use of nuclear
wave functions generated with the use of
nuclear interactions that represent
all the irreducible $A$-nucleon diagrams
up to $\nu$-th order.
Thus a transition matrix in nuclear EFT
is given by
\be
{\cal M}_{fi}^{\mbox{\tiny{EFT}}}\,=\,
<\!\Psi_{\!f}^{\mbox{\tiny{EFT}}}\,|
{\cal T}^{\mbox{\tiny{EFT}}}
\,|\Psi_i^{\mbox{\tiny{EFT}}}\!>
= <\!\Psi_f^{\mbox{\tiny{EFT}}}|
\sum_i^A{\cal O}_i^{\mbox{\tiny{EFT}}}
+\sum_{i<j}^{A}{\cal O}_{ij}^{\mbox{\tiny{EFT}}}+\,\cdots
|\Psi_i^{\mbox{\tiny{EFT}}}\!\!>\,,\label{ME-EFT}
\ee
where the superscript, ``EFT",
means that the relevant quantities are calculated
according to EFT as described above.
If this program is carried out rigorously,
it would constitute an {\it ab initio} calculation.
Let us call it ``AB-EFT".
It is to be noted that in EFT we know exactly
at what chiral order three-body operators
start to contribute to ${\cal T}$,
and that, to chiral orders
relevant to most applications
described below, there is no need for
three-body operators.
With this understanding,
we have kept only one- and two-body operators
in eq.(\ref{ME-EFT}).
The unambiguous classification of
transition operators according to their chiral orders
is a great advantage of EFT,
which is missing in eq.(\ref{ME-SNPA}).

We should point out that there exists
an alternative form of nuclear EFT
called the PDS scheme,
proposed by Kaplan, Savage and Wise~\cite{ksw},
which uses a counting scheme (often called Q-counting)
different from the Weinberg scheme.
Although many important results have been obtained
in the PDS scheme (for a review, see \eg
\cite{beaetal01}),
we concentrate here primarily on the Weinberg scheme.

We also remark that,
for very low energy nuclear phenomena whose typical
energy-momentum scale is $Q \ll m_\pi$,
even the pions can be treated
as ``heavy" particles and can be eliminated
from ${\cal L}_{{\rm eff}}$,
with the choice of
$\Lambda$ $\approx$ $m_{\pi}$.
This leads to pionless EFT~\cite{ksw,BEDetal,CRS99}.
Here there is no manifest chiral symmetry
since there is no explicit pion.
In this regime the $NN$ interactions and electroweak currents
are described by point-like contact terms.

\vspace{4mm}
\noin
$\bullet$ \hspace{3mm} {\bf Hybrid EFT}

In the above
we emphasized the formal merits of nuclear EFT.
In actual calculations, however,
we need to consider the following two points.
First, it is still a big challenge
to generate, strictly within the EFT framework,
nuclear wave functions the accuracy of which
is comparable to that of SNPA wave functions.
Secondly, whereas a chiral Lagrangian, ${\cal L}_{{\rm eff}}$,
is definite only when the values of
all the relevant LECs are fixed,
this condition cannot be readily satisfied
in many cases.
A pragmatic solution to the first
difficulty~\cite{pkmr98a}-\cite{PMS-pphep}
is to use in eq.(\ref{ME-EFT}) wave functions
obtained in SNPA;
this eclectic approach is called hybrid EFT.
A nuclear transition matrix element in hybrid EFT
is given by
\be
{\cal M}_{fi}^{hyb-EFT}\,=\,
<\!\Psi_{\!f}^{SNPA}\,|\sum_\ell^A{\cal O}_\ell^{EFT}
+\sum_{\ell<m}^A{\cal O}_{\ell m}^{EFT}
\,|\Psi_i^{SNPA}\!>\,, \label{ME-hybrid}
\ee
Because, as mentioned, the NN interactions
that generate SNPA wave functions
reproduce accurately the two-nucleon data,
the adoption of eq.(\ref{ME-hybrid})
is almost equivalent to using the empirical data themselves
to control the initial and final nuclear wave functions.
From a purely formal point of view,
hybrid EFT may be deemed as a ``regression".
But, if our goal is to obtain
a transition matrix element as accurately as possible
with the maximum help of available empirical input,
hybrid EFT does have a legitimate status.
It offers more flexibility and predictive power
than the AB-EFT, and it can be highly accurate
so long as the off-shell problem
and the contributions of three-body (and higher-body)
terms are properly addressed.
The issue of possible unknown LECs
will be discussed in connection with MEEFT
in the next subsection.

\vspace{4mm}
\noin
$\bullet$ \hspace{3mm} {\bf MEEFT}

A great merit of hybrid EFT is that
it can be used for complex nuclei ($A$~=~3, 4, ...)
with essentially the same accuracy and ease
as for the $A$=2 system.\footnote{
Here we are ignoring ``purely technical" complications
that can grow in actual numerical calculations
for higher-$A$ systems.}
We reemphasize in this connection
that the contributions of transition operators
involving three or more nucleons,
which can in principle contribute
in $A$-nucleon systems ($A \!\ge \!3$),
are intrinsically suppressed
according to chiral counting,
and hence, up to a certain chiral order,
a transition operator in an $A$-nucleon system
consists of the same EFT-based
1-body and 2-body terms
as used for the two-nucleon system.
Then, since SNPA provides high-quality wave functions
for the $A$-nucleon system,
one can calculate ${\cal M}_{fi}^{hyb-EFT}$
with precision comparable to that for
the corresponding two-nucleon case.

In most cases the one-body
operator, ${\cal O}_\ell^{EFT}$,
is free from unknown LECs.  So we may
concentrate on the two-body operator,
${\cal O}_{\ell m}^{EFT}$.
Suppose that ${\cal O}_{\ell m}^{EFT}$
under consideration contains an LEC (call it $\kappa$)
that cannot be determined with the use of
$A$=2 data alone.
It can happen that
an observable (call it $\Omega$)
in a A-body system ($A\!\ge\!3$)
is sensitive to $\kappa$ and
that the experimental value of $\Omega$
is known with sufficient accuracy.
We then can determine $\kappa$
by calculating ${\cal M}_{fi}^{hyb-EFT}$
responsible for $\Omega$
and adjusting $\kappa$ to reproduce
the experimental value of $\Omega$.
Once $\kappa$ is determined this way,
we can make {\it predictions} for any other
observables for any other nuclear systems
that are controlled by the same transition
operators.
When hybrid EFT is used in this manner,
it is called {\it more effective}
effective field theory
(MEEFT)~\cite{pkmr01,PMS-pphep}.\footnote{
It is also called EFT*.
MEEFT should be distinguished from
an earlier naive hybrid EFT model
wherein the short-range terms were
dropped altogether using an intuitive argument
based on short-range NN repulsion.}

MEEFT is a powerful formalism
for correlating various observables in different
nuclei, using the transition operators
controlled by EFT.
A further practical advantage of MEEFT
is that, since correlating the observables
in neighboring nuclei is expected to serve as
an additional renormalization,
the possible effects of higher chiral order terms
and/or off-shell ambiguities
can be significantly suppressed
with the use of MEEFT.
We will come back to this later,
when we discuss concrete examples.

\section{Examples}

\noin
$\bullet$ \hspace{2mm}
{\bf $pp$-fusion, $\nu d$ reactions and $\mu d$ capture}

A common feature of these reactions is that
a precise knowledge of the Gamow-Teller (GT)
transition matrix elements is required
in estimating their cross sections.
We show here, following
Refs.~\cite{PMS-pphep},
that the idea of MEEFT can be used very neatly
for this group of reactions.
The 1-body IA operators for the GT transition
can be fixed unambiguously from the available
1-body data.
As for the 2-body operators,
to next-to-next-to-next-to-leading order (N$^3$LO)
in chiral counting,
there appears one unknown LEC that at present
cannot be determined from data for the $A$=2 systems.
This unknown LEC,
denoted by $\widehat{d}_R$ in \cite{pkmr98b},
parametrizes the strength
of contact-type four-nucleon coupling
to the axial current.
Park {\it et al.\ }\cite{PMS-pphep}
noticed that the same LEC, $\widehat{d}_R$,
also features as the only unknown parameter
in the calculation of the tritium $\beta$-decay rate
$\Gamma_{\beta}^t$,
and they proposed to use MEEFT
to place a constraint on $\widehat{d}_R$
from the experimental information on $\Gamma_{\beta}^t$.
Since the empirical value
of $\Gamma_{\beta}^t$ is known with high precision,
and since the accurate wave functions of
$^3$H and $^3$He are available
from SNPA~\cite{Metal},
we can determine $\widehat{d}_R$
with sufficient accuracy for our purposes.
Once the value of $\widehat{d}_R$ is determined this way,
we can carry out parameter-free MEEFT calculations
for $pp$-fusion~\cite{PMS-pphep},
$\nu$-$d$ reactions~\cite{andetal-nud},
and $\mu$-$d$ capture.
(An on-going attempt to deduce $\widehat{d}_R$
from $\mu$-$d$ capture will be discussed later.)

We should mention the important role of momentum cutoff
in MEEFT.
As emphasized before, the effective Lagrangian
${\cal L}_{eff}$ is, by construction, valid
only below the specified cutoff scale $\Lambda$.
Needless to say, this basic constraint should
be respected in our nuclear EFT calculations,
and for that
we must make sure that nuclear intermediate states
involved in the computation of eq.(\ref{ME-EFT})
do not get out of this constrained world.
Since, however, the use of $V^{\rm phenom}$
introduces high momentum components above
$\Lambda_{\rm QCD}$,
we need to introduce momentum cutoff at
$\Lambda_{NN}$ to eliminate high-momentum components.
It is reasonable to implement this constraint
by requiring that the two-nucleon relative momentum
should be smaller than $\Lambda_{NN}$.
Park {\it et al.\ }\cite{PMS-pphep}
used a Gaussian cutoff function
proportional to
$\exp(-\vec{p}^2/\Lambda_{NN}^2)$
but its detailed form should not be too important.
As a reasonable range of the value of $\Lambda_{NN}$
we may choose:
500 MeV $\lsim \Lambda_{NN} \lsim$ 800 MeV,
where the lower bound is dictated by
the requirement that $\Lambda_{NN}$
should be sufficiently larger than
the pion mass (to fully accommodate pion physics),
while the upper bound reflects the fact
that our EFT is devoid of the $\rho$ meson.
Mismatch between $\Psi^{\mbox{\tiny{EFT}}}$ and
$\Psi^{\mbox{\tiny{SNPA}}}$
only affects short-distance behavior,
and we expect that $\Lambda_{NN}$-independence
is a measure of model independence
of an MEEFT calculation.

For a given value of $\Lambda_{NN}$ within the above range,
$\widehat{d}_R$ is tuned to reproduce
$\Gamma_{\beta}^t$,
and then the rates for pp-fusion,
the $\nu d$ reactions and $\mu d$ capture
are calculated.  
Before giving an account of
the individual results,
we should point out a notable common feature.
Although the optimal value of $\widehat{d}_R$
varies significantly as a function of $\Lambda_{NN}$,
the observables (in our case the above three reaction
rates) exhibit remarkable stability
against the variation of $\Lambda_{NN}$
(within the above-discussed physically reasonable range).
This stability may be taken as an indication
that the use of MEEFT for
inter-correlating the observables in neighboring nuclei
{\it effectively} renormalizes various effects,
such as the contributions of higher-chiral order terms,
mismatch between the SNPA and EFT wave functions, etc.
This stability is essential in order
for MEEFT to maintain its predictive power.

\subsection{The $pp$-fusion S-factor}

The $S$-factor is related
to the cross section $\sigma$ as
$$
\sigma(E)=\frac{S(E)}{E}
\exp \{ -2\pi \eta(E)\}\,,
$$
where $E$ is the incident energy
in the center-of-mass system, and
$\eta(E)\equiv\frac{Z_1Z_2e^2}{\hbar v}$
is the Coulomb penetration factor.
The $S$-factor for $pp$-fusion at threshold
is related to the transition matrix element ${\cal M}$
as~\cite{ADEetal2010}
\be
S_{pp}(0) =6\pi^2m_p\,\alpha\, {\rm ln}2\,
\frac{1}{\gamma^3}(g_A/g_V)^2
\frac{f_{pp}^R}{(ft)_{0^+\to0^+}}
|{\cal M}|^2 \label{SppM}
\ee
where $\gamma=(2\mu B_d)^{1/2}$ = 0.2316 fm$^{-1}$
is the deuteron binding wave number,
and ${f_{pp}^R}$ is the phase space factor
including radiative corrections.
As mentioned,
the solar model and stellar evolution theory require
1\% precision in $S_{pp}$~\cite{ADEetal2010}.

It is useful to first recall what was achieved in SNPA.
A benchmark calculation of $S_{pp}$ in SNPA was carried out
by Schiavilla {\it et al.}~\cite{schetal98}.
These authors took full advantage
of the fact, first pointed out in Ref.~\cite{CARetal91},
that the Gamow-Teller (GT) matrix elements
for $pp$-fusion and tritium $\beta$-decay
are closely correlated.
For each of the two-body GT transition operators
with different $r$-dependences
($r$ = relative distance of the two nucleons),
the transition density for $pp$-fusion
is found to be accurately proportional
to that for tritium $\beta$-decay.
This means that a single multiplicative constant
can match them all.
Schiavilla {\it et al.}~\cite{schetal98} fine-tuned
the $N\Delta$-axial coupling constant,
which featured in a dominant two-body axial current
diagram ($\Delta$-particle excitation diagram),
and which was only loosely controlled from other
sources of information.
After this fine-tuning, $S_{pp}$ was calculated
with estimated uncertainties of $\sim$0.3 \%;
here the error estimate came from the range of variation
in $S_{pp}(0)$ for the five different
high-precision phenomenological potentials used in
\cite{schetal98}.  A further examination
of this error estimation will be described below.

Park {\it et al.}~\cite{PMS-pphep}
carried out an MEEFT calculation of $S_{pp}(0)$
and obtained
\be
S_{pp}(0)=3.94\!\times\!(1\pm0.008)
\times 10^{-25}\,{\rm MeV\, b}\,.\label{Spp}
\ee
$S_{pp}(0)$ is found to change
only by $\sim$0.1\%
against changes in $\Lambda_{NN}$,
ensuring the robustness of the MEEFT calculation
in this respect.
The convergence pattern in chiral expansion
for the $pp$-fusion
transition amplitude is reported to be~\cite{PMS-pphep}:
${\cal M} \sim (1+0+0.1\%+0.9\%+0.4\%)$.
It is noted that the 2nd and 3rd order terms
are accidentally small,
but that the 4th and 5th order are roughly in
conformity with expansion in
$m_\pi/\Lambda_\chi\approx \frac{1}{7}$.
The 3-body currents contribute at the sixth order.
Uncertainties due to the
higher order contributions estimated
from the size of the 5th-order contribution
are 0.4 \% (in amplitude), or
0.8 \% (in probability).\footnote{In \cite{PMS-pphep},
the relative error of 0.4 \% was assigned
to $S_{pp}(0)$.  However, it has been found subsequently
that the 0.4 \% uncertainty should have been assigned to
the transition amplitude, not to the transition probability;
we therefore double the error estimate quoted in \cite{PMS-pphep}
and adopt 0.008 in eq.(\ref{Spp}). }
The MEEFT result, eq.(\ref{Spp}),
is consistent with that obtained in SNPA~\cite{schetal98}.
Meanwhile, the fact that MEEFT allows us
to make error estimation is a notable advantage
over SNPA.
It is worth emphasizing that the accuracy
achieved in eq.(\ref{Spp}) represents an improvement
by a factor of $\sim$5
over the previous results
based on a naive hybrid EFT~\cite{pkmr98b}.

The calculation of $S_{pp}$ in pionless EFT
was performed up to NNLO by Kong and Ravndal~\cite{KR01},
and up to fifth order by Butler and Chen~\cite{BC01};
see also Ando {\it et al.}~\cite{ANDetal08}.
Here again, up to the order considered,
there is only one unknown LEC, denoted by $L_{1,A}$,
which (analogously to $\widehat{d}^R$) parameterizes
the strength of four-nucleon contact axial-vector coupling.
The results of Refs.~\cite{KR01,BC01} are:
(i) the central value of $S_{pp}(0)$
is consistent with those obtained in SNPA and MEEFT;
(ii) the theoretical uncertainty in $S_{pp}(0)$
is $\sim$3 \%, which is significantly larger than
the results obtained in SNPA and MEEFT.
One thing to be noted here is that the PDS scheme~\cite{ksw}
(used in pionless EFT)
adopts an expansion scheme
for transition amplitudes themselves,
without employing the concepts of the potential
or wave functions.
In certain contexts this feature may be an advantage,
but its disadvantage in the present context
is that one cannot readily relate
the transition matrix elements
for an $A$-nucleon system with those for
the neighboring nuclei;
in PDS, each nuclear system requires a separate
parameterization.
Therefore, $L_{1A}$ in pionless EFT
needs to be determined from the empirical
value of a 2-body observable, which however
is at present available only with large experimental error.
This situation is reflected in the rather large 3 \%
uncertainty in $S_{pp}(0)$ calculated in pionless EFT.

In a very recent review~\cite{ADEetal2010},
there has been an attempt to deduce
the most updated estimate for $S_{pp}(0)$,
and the recommended value is
\be
S_{pp}(0)=  4.01\times(1\pm 0.009)\times10^{-24}
\;\;{\rm MeV\, b}\label{SppRMP}.
\ee
This estimate arises as follows.
In eq.(\ref{SppM}), the matrix element ${\cal M}$
is assigned the value corresponding to the central value
obtained in SNPA~\cite{schetal98}
and MEEFT~\cite{PMS-pphep};
these two approaches give exactly the same value.
For the other inputs that enter into eq.(\ref{SppM}),
the most updated values are used;
some of them slightly differ from the values
used in the earlier publications.
The relative uncertainty of 0.009 in eq.(\ref{SppRMP})
is dominated by the 0.8 \% uncertainty in $|{\cal M}|^2$
estimated from the MEEFT result~\cite{PMS-pphep}
and the 0.5 \% uncertainty coming from the experimental
error in the $nn$ scattering length $a_{nn}$.
These uncertainties and other smaller uncertainties
are added in quadrature.
Eq.(\ref{SppRMP}) indicates that $S_{pp}(0)$
now can be estimated within 1\% precision.

\subsection{$\nu$-$d$ reactions}

As mentioned, $\nu$-$d$ reactions are
important in connection with the SNO experiments
on neutrino oscillations.
Within SNPA a detailed calculation of
the $\nu$-$d$ cross sections, $\sigma(\nu d)$,
was carried out in Refs.~\cite{kn94,NSGK,NETAL}.
Nakamura {\it et al.}~\cite{NETAL},
following Schiavilla {\it et al.}'s approach~\cite{schetal98},
used the two-body current whose strength was fine-tuned
to reproduce the tritium $\beta$-decay rate.
With the use of this fine-tuned two-body exchange current,
Nakamura {\it et al.} calculated $\sigma(\nu d)$
for four types of reactions:
charged-current and neutral-current reactions on $d$
for both neutrino and anti-neutrino cases.
Meanwhile, Butler, Chen and Kong (BCK)~\cite{EFT}
carried out a pionless-EFT calculation of
the $\nu$-$d$ cross sections.
The results obtained agree with those of
Nakamura {\it et al.}~\cite{NETAL}
in the following sense.
A pionless-EFT calculation,
as mentioned in the discussion of $pp$-fusion,
involves one unknown LEC, $L_{\rm 1A}$,
which represents the strength of
a four-nucleon axial-current coupling term.
BCK determined $L_{\rm 1A}$ by requiring that
the $\nu d$ cross sections of NETAL be reproduced
by their pionless-EFT calculation.
With the value of $L_{\rm 1A}$
fine-tuned this way,
the $\sigma(\nu d)$'s obtained by BCK
show a perfect agreement with
those of Nakamura {\it et al.} for all the four reactions
mentioned above, and for the entire solar neutrino energy range,
$E_\nu\lsim$ 20 MeV.
Moreover, the optimal value,
$L_{\rm 1A}=5.6\,{\rm fm}^3$, found
by BCK~\cite{EFT} is consistent
with the order of magnitude of $L_{\rm 1A}$
expected from the naturalness argument
(based on a dimensional analysis),
$|L_{\rm 1A}|\le 6\,{\rm fm}^3$.
The fact that an EFT calculation
(with one parameter fine-tuned)
reproduces the results of SNPA very well
strongly suggests the robustness of
the SNPA results for $\sigma(\nu d)$.

Even though it is reassuring
that the $\nu$-$d$ cross sections
calculated in SNPA and pionless-EFT
agree with each other
(in the sense explained above),
it is desirable to carry out an EFT calculation
that is free from any adjustable LEC.
Fortunately, MEEFT allows us
to carry out an EFT-controlled parameter-free calculation
of the $\nu$-$d$ cross sections,
because the only LEC appearing therein, $\widehat{d}_R$,
has already been determined from $\Gamma_\beta^t$~\cite{PMS-pphep}.
Ando {\it et al.}~\cite{andetal-nud}
took advantage of this situation
and carried out an MEEFT calculation of $\sigma(\nu d)$.
The results were found to agree within 1\% with
$\sigma(\nu d)$'s obtained in SNPA~\cite{NETAL}.
These results show
that the $\nu$-$d$ cross sections used
in interpreting the SNO experiments~\cite{ahmetal}
are reliable at the 1\% precision level.

{There have been subsequent attempts
to calibrate $L_{1A}$ from reactor
$\bar{\nu}$-deuteron breakup reactions~\cite{BCV02},
and also from a ``self-calibrating" analysis
of the SNO data~\cite{chr03}.}
The $\sigma(\nu d)$'s
corresponding to the range of $L_{1A}$
obtained in these analyses
are consistent with the $\sigma(\nu d)$'s obtained
in SNPA and MEEFT.
However, the resulting constraints on $L_{1A}$
are not stringent.
We have mentioned that both $L_{1A}$
and $\widehat{d}_R$ represent the strength
of axial-current-four-nucleon contact coupling.
It is to be noted, however, that
$L_{1A}$ belongs to pion-less EFT,
while $\widehat{d}_R$ to pion-ful EFT.
In the pion-ful EFT, because of the strong tensor force,
the exchange current involving the deuteron $d$-state
is important, and the s-wave exchange current
arising from the $\widehat{d}_R$ term is separate from
this tensor-force effect.
By contrast, in the pion-less EFT,
the explicit $d$-wave term is a higher-order correction,
and hence the $s$-wave $L_{1A}$ term must subsume
the strong tensor-force contributions.
It would be informative to investigate the relation
between $L_{1A}$ and $\widehat{d}_R$ from this perspective.

\noin
$\bullet$\hspace{2mm}{\bf Off-shell effects}

In introducing hybrid EFT, we have replaced
$|\Psi\!>^{EFT}$ for the initial
and final nuclear states in eq.(\ref{ME-EFT})
with the corresponding $|\Psi\!>^{SNPA}$;
see eq.(\ref{ME-hybrid}).
This replacement may bring in
a certain degree of model dependence,
called the off-shell effect, because
the phenomenological NN interactions
are constrained only by the
on-shell two-nucleon observables.\footnote{
In a consistent theory, physical observables
are independent of field transformations
that lead to different off-shell behaviors,
and therefore the so-called off-shell effect
is not really a physical effect.
In an approximate theory, observables may
exhibit superficial dependence on off-shell behavior,
and it is customary to refer to this dependence
as an off-shell effect.}
This off-shell effect, however, is expected
to be small for the reactions under consideration,
since they involve low momentum transfers
and hence are not extremely sensitive to
the short-range behavior of the nuclear wave functions.
One way to quantify this expectation is to compare
a two-nucleon relative wave function generated by the phenomenological
potential with that generated by an EFT-motivated potential.
Phillips and Cohen~\cite{pc00} made such a comparison
in their analysis of the 1-body operators
responsible for electron-deuteron Compton scattering,
and showed that a hybrid EFT should work well up to
momentum transfer 700 MeV.
A similar conclusion is expected to hold
for a two-body operator,
so long as its radial behavior
is duly ``smeared-out" reflecting
a finite momentum cutoff.
Thus, hybrid EFT as applied to low energy
phenomena is expected to be practically
free from the off-shell ambiguities.
The off-shell effect should be even less significant
in MEEFT, wherein an additional ``effective" renormalization
is likely to be at work (see subsection 3.1).

We now discuss briefly another interesting development,
due to Tom Kuo and his colleagues~\cite{kuo},
which can shed much light
on the reliability of a hybrid EFT or MEEFT calculation.
As mentioned, a ``realistic phenomenological" nuclear
interaction, $V_{ij}$ in eq.(\ref{HSNPA}),
is determined by solving the Schr\"{o}dinger equation,
eq.(\ref{Sch}), for the $A$=2 system and fitting the results
to the full set of two-nucleon data
up to the pion production threshold energy.
So, physically, $V_{ij}$ should reside in a momentum regime
below a certain cutoff, $\Lambda_c$.
In the conventional treatment, however,
the existence of this cutoff scale is ignored,
and eq.(\ref{Sch}) is solved,
allowing the entire
momentum range to participate.
Kuo {\it et al.\ }proposed to construct an
{\it effective low-momentum} potential
$V_{low-k}$ by eliminating
(or integrating out) from $V_{ij}$
the momentum components higher than
$\Lambda_c$, and calculated $V_{low-k}$'s
corresponding to many well-established examples
of $V_{ij}$'s.
Remarkably, it was found that all these $V_{low-k}$'s give
identical results for the half-off-shell T-matrices,
even though the ways short-range physics is encoded
in these $V_{ij}$'s are highly diverse.
This implies that the $V_{low-k}$'s are free from
the off-shell ambiguities, and therefore the use of
$V_{low-k}$'s is essentially equivalent to employing
$V_{ij}^{EFT}$ (that appeared in eq.(\ref{HEFT})),
which by construction should be model-independent.
Now, as mentioned, our MEEFT calculation has
a momentum-cutoff regulator, and this essentially
ensures that the matrix element,
${\cal M}_{fi}^{hyb-EFT}$, in eq.(\ref{ME-hybrid})
is only sensitive to the half-off-shell T-matrices
that are controlled by $V_{low-k}$ instead of $V_{ij}$.
Therefore, we can expect that the MEEFT results
reported here are essentially free from the off-shell
ambiguities.

Recently, Mosconi {\it et al.}~\cite{MOSetal07}
have compared $\nu d$ cross sections calculated
for various models that differ in the way
the ``nuclear PCAC requirement" is implemented,
and concluded from this comparison that
the $\sigma(\nu d)$'s obtained in SNPA, MEEFT
and pionless EFT have uncertainties as large as 2-3 \%.
Mosconi {\it et al.} have reached this conclusion by examining
the scatter of unconstrained calculations of $\sigma(\nu d)$.
However, all state-of-the-art calculations use
$\Gamma^t_\beta$ to reduce
two-body current and other uncertainties.
Once this constraint is imposed,
the scatter in the calculated values of $\sigma(\nu d)$
should be significantly reduced.

\subsection{$\mu d$ capture and MuSun experiment}

Although it is likely
that the determination of $\widehat{d}_R$
from $\Gamma_{\beta}^t$ is good enough
for practical purposes,
it is worthwhile to study
a possibility to fix $\widehat{d}_R$ with the use
of an observable belonging to the $A$=2 systems.
A promising candidate is the $\mu d$-capture process,
$\mu^-+d\rightarrow \nu_\mu+n+n$.
Even though rather large energy-momentum transfers
caused by the disappearance of a $\mu^-$
seem to make the applicability of EFT here a delicate issue,
we can show~\cite{APMK-muD} that,
as far as the hadron sector is concerned,
$\mu$-$d$ capture is in fact a reasonably ``gentle" process.
This is because: (1) the $\nu_\mu$
carries away most of the energy,
and (2) there is a large enhancement
of the transition amplitude
in a kinematic region where the relative motion
of the final two nucleons is low enough to
justify the use of EFT.
Thus $\mu d$ capture can be used for
controlling $\widehat{d}_R$,
if the quality of experimental data
improves sufficiently.
The MuSun experiment at PSI
(proposed in 2008)~\cite{MuSun}
aims to determine the $\mu d$ capture rate $\Gamma_{\mu d}$
with $\sim$1.5 \% accuracy;
for a review, see \cite{KK2010}.
This will enable us to determine $\widehat{d}^R$
within $A$=2 system.
In this connection, it is important to point out
that $g_P\equiv F_A(q^2=-0.88 \,m_\mu^2)$
is totally under control by now.
Previously, the main goal of $\mu$-nucleus capture experiments
was to obtain information on $g_P$ given the fact that
$\mu$ capture goes much faster on nuclei (with high $Z$)
than on a proton.
This goal, however, has been perennially elusive
because of nuclear physics complications.
It is therefore highly noteworthy
that a HB$\chi$PT calculation~\cite{BKM94}
gives a robust prediction, $g_P=8.26\pm 0.23$,
and that this value is in good agreement
with the experimental value, $g_P=7.7\pm 1.1$,
obtained in the MuCap Experiment at PSI~\cite{ANDetal07}.
The fact that the 1-body current is well known
allows us to concentrate on the 2-body part.

Recent SNPA calculations of $\mu$-$d$ capture
can be found in Refs.~\cite{TKK,ADAetal90,RICetal2010};
According to Tatara {\it et al.}~\cite{TKK},
the hyperfine-doublet capture rate
$\Gamma_d$ (by far the dominant capture channel)
is $\Gamma_d=397-400 \,{\rm s}^{-1}$,
and the results are stable against the use of different
$NN$ potentials available at the time of their work.
The latest calculation by Ricci {\it et al.}~\cite{RICetal2010}
reports $\Gamma_d= 416 \,{\rm s}^{-1}$,
and that there is non-negligible $NN$-potential dependence
between the earlier and new-generation realistic phenomenological
potentials.
Thus the results obtained within SNPA exhibit some scatter,
and although Ricci {\it et al.} offer rather detailed discussion
of its possible origin, further investigation
seems needed.
An MEEFT calculation of $\Gamma_d$ was
carried out by Ando {\it et al.}~\cite{APMK-muD},
yielding the value $\Gamma_d=386\,{\rm s}^{-1}$
with a high degree of stability against variations
in $\Lambda_{NN}$.
A pionless EFT calculation was performed by
Chen {\it et al.}~\cite{CIJL05} but this approach
is reliable only for limited kinematical regions,
because of the very low momentum cutoff
inherent with pionless EFT.

Comparison with experiment is complicated
by the existence of two barely overlapping
experimental results:
$\Gamma_d^{\rm exp}=470\pm29\,{\rm s}^{-1}$~\cite{BARetal86},
and
$\Gamma_d^{\rm exp}=409\pm40\,{\rm s}^{-1}$~\cite{CARetal89}.
All the above-mentioned theoretical values are
consistent with the value of $\Gamma_d^{\rm exp}$
reported in \cite{CARetal89},
and disagree with that given in \cite{BARetal86}.

Since the MuSun experiment
envisages to measure $\Gamma_d$
with 1.5 \% precision,
it is important to match the accuracy of
theoretical prediction.
Most recently, a highly elaborate calculation
of $\Gamma_d$ has been performed
in both SNPA and MEEFT by
Marcucci {\it et al.}~\cite{MARetal2010}\footnote{
The authors of this work use the term EFT*
instead of MEEFT.}.
The nuclear wave functions are generated from
the Argonne $V_{18}$ potential (for SNPA)
or from a chiral N3LO two-nucleon
potential~\cite{EM03} (for MEEFT).
As in the pp-fusion calculation,
one parameter needs to be fixed
(the $N$-to-$\Delta$ axial coupling constant
in SNPA or $\widehat{d}_R$ in MEEFT).
This is fixed by reproducing
the tritium $\beta$-decay rate.
The model dependence with respect to the adopted interactions
and current (including cutoff dependence in the case of MEEFT)
is found to be weak.
Marcucci {\it et al.} give
\be
\Gamma_d=392\pm2.3\,{\rm s}^{-1}\,.\label{GDMar}
\ee
It is interesting to note that this result is close
to that obtained by Ando {\it et al.}~\cite{APMK-muD}.
We also remark that what Marcucci {\it et al.} refer to
as MEEFT is actually ``less hybrid" in nature
than the MEEFT used by Ando {\it et al.} in that the former
employs the chiral N3LO two-nucleon potential
instead of a phenomenological $NN$ potential.
When the MuSun result becomes available,
one can use it in two ways -- compare it with
the theoretical prediction in eq.(\ref{GDMar})
based on $\Gamma^t_\beta$,
or use it to determine $\widehat{d}_R$
from the $A$=2 system
(without referring to the $A$=3 system).

\subsection{Hep and Hen reactions}

\noin
$\bullet$\hspace{3mm} {\bf Hep reaction}

As mentioned, the Hep reaction,
$^3{\rm He}+p \to\, \!^4{\rm He}+\nu_e + e^+ $,
produces solar neutrinos
of the highest maximum energy
($E_\nu^{\rm max}$ = 18.8 MeV),
although the flux is extremely low.
An accurate estimation of this cross section
is a particularly challenging task because:
(1) the contribution of the leading-order
1-body GT operator is highly suppressed
due to the approximate wave function
orthogonality~\cite{WB67},
and (2) there is a strong cancellation between
the 1-body and 2-body GT matrix elements
\cite{CARetal91,Metal}.
Reflecting this situation, estimates of $S_{\rm Hep}(0)$
changed by orders of magnitude
from Salpeter's original value,
$S_{\rm Hep}(0)_{\rm Salpeter}=
630\times 10^{-20}$ keV-b
to modern values;
{for a review, see \cite{KP04}}.
Park {\it et al.}~\cite{PMS-pphep}
carried out an MEEFT calculation of $S_{\rm Hep}(0)$
with the use of $\widehat{d}_R$
determined from $\Gamma^t_\beta$
(the same procedure as used in calculating
$S_{pp}(0)$), and obtained
\be
S_{Hep}(0)= (8.6\pm1.3)\times10^{-20}\,
{\rm keV\,b}\,,\label{SHep}
\ee
where the error spans the range of
the $\Lambda_{NN}$ dependence for
$\Lambda_{NN}$ = 500 - 800 MeV.
Again, the MEEFT result agrees with that
obtained in SNPA by Marcucci {\it et al.\ }~\cite{Metal}:
$S_{\rm Hep}(0)_{\rm SNPA}=9.6 \times 10^{-20}$ keV-b.
The above-mentioned large cancellation between
the 1-body and 2-body contributions in this case
amplifies the cutoff dependence of
$S_{Hep}(0)$, but the error quoted in
eq.(\ref{SHep}) is still small enough
for the purpose of analyzing the existing
Super-Kamiokande data~\cite{SuperK-Hep}.

\newpage
\noin
$\bullet$\hspace{3mm} {\bf Hen reaction}

The Hen reaction,
$^3{\rm He}+\!n \to \,^4{\rm He}+\!\gamma $,
is very similar to Hep except, of course,
the weak interaction is replaced by the EM interaction.
Thus, just like the Gamow-Teller matrix element for Hep,
the leading-order 1-body M1 matrix element for Hen
is highly suppressed due to the pseudo-orthogonality
of the initial and final nuclear wave functions.
Moreover, there is strong cancellation
between the hindered 1-body contribution
and the 2-body exchange-current contribution,
leading to a further suppression of the Hen reaction rate.
Thus, like Hep, Hen is sensitive to higher order
terms in chiral expansion.
Meanwhile, the experimental value of $\sigma_{\rm Hen}$
is known rather accurately:
$\sigma_{\rm Hen}^{\rm exp}=
(54\pm6)$ $\mu b$~\cite{WOLetal89},
or
$\sigma_{\rm Hen}^{\rm exp}=
(55\pm3)$ $\mu b$~\cite{WERetal91}.
Therefore, Hen provides a good testing ground
for the validity of MEEFT as used for Hep.
A very elaborate MEEFT calculation of Hen
was recently carried out by
Lazauskas, Song and Park~\cite{LSP09b}.
Up to N$^3$LO considered by these authors,
the M1 operators contain two LECs,
{which can be determined from}
the magnetic moments, $\mu(^3{\rm H})$
and $\mu(^3{\rm He})$.
The initial and final nuclear wave functions
are obtained from the rigorous Faddev-Yakubowski equations
\cite{Yakub} for five sets
of high-precision phenomenological potentials.
Lazaukas {\it et al.}'s calculation gives
\be
\sigma_{\rm Hen}= (38 \sim 58) \,\mu b\,,
\label{HenLSP}
\ee
with high stability against changes in $\Lambda_{NN}$
in the range 500 $\sim$ 900 MeV.
The authors emphasized the importance
of long-range nuclear correlation effects
in obtaining an accurate estimate of $\sigma_{\rm Hen}$.
To explain this point,
it is useful to first discuss what
Song {\it et al.}~\cite{SLP09}
pointed out in their MEEFT study
of the M1 properties of the $A$=3 nuclei.
They noticed that the M1 matrix elements (MEs)
in the $A$=3 nuclei calculated in MEEFT
for various realistic potentials
have strong correlation
with the triton binding energies $B_3$
calculated with the corresponding realistic potentials.
In fact, these quantities fall on a  well-defined single curve.
Meanwhile, since $B_3$ governs
the long distance contribution to the MEs,
the model dependence (that is, variations in MEs
calculated for nuclear potentials
that give different values for $B_3$)
cannot be lifted by renormalizing
the local (or short-range) operators.
However, the use of the empirically found
correlation curve between the MEs and $B_3$ allows
us to drastically reduce scatters in the calculated MEs.
This is achieved by introducing the constraint
that one should accept only those values of the MEs
which, along the correlation curve, have values of $B_3$
consistent with its experimental value.
This constraint is found to essentially
eliminate the model dependence in the MEs~\cite{SLP09}.
Now, Lazauskas {\it et al.}~\cite{LSP09b}
noticed that a very similar feature exists also for Hen.
Defining the quantity $\zeta$ by
$\zeta\equiv[q(a_{nHe^3}/r_{He^4}^2]^{-2.75}$
($q$ = photon momentum),
Lazauskas {\it et al.} found that $\sigma_{\rm Hen}$
and $\zeta$ are so well correlated that
the $\sigma_{\rm Hen}$-$\zeta$ plot
forms a sharply defined straight line.
Then, again, by accepting only those values of
$\sigma_{\rm Hen}$ which are consistent
with the empirical value of $\zeta$,
one can deduce the model dependence
of $\sigma_{\rm Hen}$ significantly.
The estimate of $\sigma_{\rm Hen}$ given
in eq.(\ref{HenLSP}) was obtained
through this method.\footnote{{
The main point here is {\it not} a linear relation
between $\sigma_{\rm Hen}$ and $\zeta$;
after all, the latter is a rather ``esoteric" quantity,
introduced just to ``dramatize" the correlation
between $\sigma_{\rm Hen}$ and binding properties
of the relevant nuclei.
The message here is that, to get a reliable estimate
of $\sigma_{\rm Hen}$, one must employ,
even among so-called high-precision nuclear potentials,
only those that can reproduce the binding properties
of the $A$=3 and 4 nuclei with needed accuracy.
This criterion does not require
the introduction of $\zeta$.}}
The fact that this theoretical value is in good agreement
with the above quoted experimental value
renders strong support for the basic soundness of
the MEEFT approach.
{We consider it noteworthy that MEEFT works well {\em also}
for a highly suppressed process
wherein higher-chiral order
terms become important; see Ref.~\cite{pkmr2000}.}
The examples of Hen and the MEs in the A=3 systems
also demonstrate that it is
profitable to avail ourselves of
the merit of EFT (to control short-range physics)
and the strength of SNPA
(to control long-range nuclear correlations),
as we do in MEEFT.

\subsection{$\mu$ -$^3{\rm He}$ capture}

Ackerbauer {\it et al.}~\cite{ACKetal08} carried out
a high-precision measurement
of the capture rate for
$\mu^-{\rm He}^3\to\nu_\mu t$.
Achieving astonishing accuracy of 0.3 \%,
they reported the experimental value
$\Gamma(\mu ^3{\rm He})^{\rm exp}=1494(4)\,{\rm s}^{-1}$.
A detailed MEEFT calculation of this capture rate
was performed by Gazit~\cite{GAZ08}.
Here again, the only unknown LEC, $\widehat{d}_R$,
was fixed by fitting to $\Gamma^t_\beta$ for each given
value of the cutoff parameter $\Lambda_{NN}$
(ranging 500 $\sim$ 800 MeV),
and then $\Gamma(\mu ^3{\rm He})$ was calculated.
The results are summarized as
$\Gamma(\mu ^3{\rm He})
=1499(2)_\Lambda(3)_{NM}(5)_t(6)_{RC}$,
where the first error comes from the
$\Lambda_{NN}$ dependence,
the second error from uncertainties in extrapolating
the nucleon weak form factors to finite momentum transfer,
and in the choice of specific nuclear models,
the third error from experimental accuracy
for $\Gamma^t_\beta$, and the fourth error from
uncertainty in radiative corrections.
The theoretical value (with total error $\sim$ 1 \%)
agrees well with $\Gamma(\mu ^3{\rm He})^{\rm exp}$,
providing a yet another case to support the validity
of MEEFT;
see also Gazit {\it et al.}~\cite{GQN09},
where a full EFT calculation for tritium $\beta$-decay
and the tritium binding energy is presented.

In Ref.~\cite{MARetal2010}, quoted earlier in connection
to $S_{pp}$ calculations, Marcucci {\it et al.} also
carried out an elaborate calculation of
$\Gamma(\mu ^3{\rm He})$ in both SNPA and MEEFT,
{ and they found excellent agreement between the two approaches.}
Combining the results coming from these two methods,
Marcucci {\it et al.} arrived
at a {highly accurate} estimate
$\Gamma(\mu ^3{\rm He})=1484\pm13\,{\rm s}^{-1}$.
{Again, the theoretical value agrees very well
(within quoted errors) with the above experimental value.}

\subsection{Neutron-neutron scattering length $a_{nn}$}

In the construction of a high-precision phenomenological
$NN$ potential, $V_{NN}^{\rm phenom}$,
the experimental errors in the $nn$ scattering length,
$a_{nn}$, are embedded in the overall $\chi^2$-value,
and it is not obvious
to what extent the existing samples
of $V_{NN}^{\rm phenom}$ directly reflect the
uncertainties in $a_{nn}$.
Meanwhile, low-energy electroweak transitions
in lightest nuclei are in general rather sensitive to
the precise value of $a_{nn}$.
For instance, in \cite{ADEetal2010},
uncertainty in the $pp$-fusion $S$-factor,$S_{pp}(0)$,
due to the existing uncertainty in $a_{nn}$
is estimated to be $0.5\%$,
which is not devastatingly large but rather significant.
Apart from this point,
it is of general importance to determine,
as accurately as possible, the empirical value of $a_{nn}$,
one of the fundamental parameters characterizing
the $NN$ interaction.
Since extraction of $a_{nn}$ from the analysis of
$nd\to nnp$ reaction at present involves some
inconsistencies~\cite{HUHetal2000,WRW06,GONetal1999},
the standard value
$a_{nn}=-18.63\pm 0.27{\rm (exp)}
\pm0.30{\rm (theor)}$ fm~\cite{CHEetal08}
is based on the $\pi^- d \to \gamma nn$ reaction.
The theoretical framework responsible for this result
uses phenomenological $NN$ interactions,
and intrinsic short-distance physics uncertainties
in this framework limit the theoretical precision.
Gardestig and Phillips~\cite{GP06,Gardestig09}
investigated the possibility of using EFT
to reduce theoretical uncertainties
in the extraction of $a_{nn}$.
They have shown that short-range physics involved
in the reaction can be parameterized using the same LEC,
$\widehat{d}_R$, as introduced in \cite{PMS-pphep},
and that, if for the sake of demonstration
the currently accepted value of $\widehat{d}_R$ is used,
the theoretical uncertainty in $a_{nn}$ can be reduced
down to $\Delta a_{nn}{(\rm theor)}$ = 0.05 fm.
This is quite an impressive reduction
{(by a factor of 6) of
the existing theoretical uncertainty.}

\subsection{Radiative corrections}

At the level of a few percent precision,
radiative corrections (RCs) become an important issue.
For instance, in $\mu p$ capture,
the electroweak RC in the semileptonic reaction
and the QED corrections to the muonic atom wave function
are found to increase the capture rate
by $2.8\pm0.4$ \%~\cite{CMS07}.
The best available estimates of RCs for
single-nucleon processes have been given
by Marciano and Sirlin~\cite{MS86}
for neutron $\beta$ decay,
and by Czarnecki {\it et al.}~\cite{CMS07}
for $\mu p$ capture.
In these works RCs of order $\alpha$ is decomposed
into so-called outer and inner corrections~\cite{Sirlin}.
The outer correction is a universal function
of the lepton energy and is independent of
the details of hadron physics, whereas the inner correction
is affected by short-distance physics and the hadron structure.
The inner corrections coming from
$\gamma$-(weak boson) loop diagrams
are divided into high-momentum and low-momentum parts.
The former is calculated in the current quark picture,
while the latter is evaluated
with the use of the phenomenological
weak form factors of the nucleon.
As regards RCs in multinucleon systems,
all existing works (except those related to the
$0^+\to0^+$ Fermi transitions),
concentrate on the single-nucleon RC diagrams
because the genuinely two-body-type RCs are expected
to be small.
Kurylov {\it et al.}~\cite{KRP03} carried out extensive studies
of RCs in the $A$=2 systems
based on the Marciano-Sirlin approach~\cite{MS86},
and their results have been widely used in analyzing
the $A$=2 weak-interaction observables
(including those discussed earlier in this article).
According to \cite{KRP03},
with the use of $G_F$ from $\mu$-decay,
and the ``effective" $g_A$ obtained from neutron
$\beta$-decay,
radiative corrections specific to $pp$-fusion
is estimated to be $\sim$ 3-4 \%
(with 0.1 \% uncertainty).

Although the existing estimates of RCs
based on the Marciano-Sirlin approach~\cite{MS86}
are believed to be reliable at the level of quoted accuracy,
it is worth noting that HB$\chi$PT offers
a model-independent framework to calculate RCs,
insofar as all the relevant LECs are known.
The first HB$\chi$PT evaluation of RCs
for neutron $\beta$ decay was carried out to order $\alpha$
by Ando {\it et al.}~\cite{ANDetal04}.
In this approach the short-distance physics contributions
can be condensed into two LECs,
one related to the Fermi constant $G_F$
and the other to the axial coupling constant $g_A$.
If one can determine these LECs from empirical data,
one would have a totally model-independent way
to evaluate RCs to neutron $\beta$ decay.
A similar situation is expected to exist for any
low-energy weak-interaction processes
on a single nucleon.
Even if one cannot determine these LECs
at present,
it seems of great interest to correlate
RCs for various weak processes on a single nucleon
taking advantage of the fact that we know exactly
what LECs are involved.

\section{Summary}

Despite the limited scope of topics covered,
we hope we have succeeded in conveying the message
that MEEFT is a highly reliable
predictive framework for computing transition amplitudes
for a large class
of electroweak processes in light nuclei.
We remark that,
in each of the cases for which both SNPA and MEEFT calculations
have been performed, it has been found that
the results of the two methods agree very well.

{Although MEEFT has been registering
and will continue to score significant successes,}
we should mention that
great progress is being made
in attempts to carry out full EFT (or ``AB-EFT") calculations
in which both the transition operators
and nuclear wave functions are derived
from HB$\chi$PT;
see. e.g., \cite{EHM09,KOEetal09,RoZetal10}.
{ We believe that, as increasingly higher order terms
are included in AB-EFT calculations,
their results will come increasingly closer to those
of MEEFT, which are anchored on accurate experimental inputs.}
Although the developments
in AB-EFT calculations are enormous and highly impressive,
we will not discuss them here;
we trust they will be covered in Ulf Meissner's
contribution to this volume.

As mentioned,
Riska and Brown's ground-breaking work
on the $np\to\gamma d$ reaction almost three decades ago
was a precursor of explosive developments
in {\em high-precision} calculations of electroweak processes
in light nuclei.
We hope Gerry is happy to see the enormous growth of the
field in which he has played such an important role.

\vspace{5mm}
\noindent {\bf Acknowledgment}

It is our great pleasure to contribute this article
to the Festschrift in honor of Gerry Brown's 85th birthday.
A detailed account of MR's life-long scientific collaboration
with Gerry is given in another article contributed to this volume.
KK started to work with Gerry Brown in 1986,
when Gerry kindly invited him to spend
a sabbatical year at Stony Brook.
KK wishes to express his deepest gratitude
to Gerry for having been an inexhaustible source
of inspiration and encouragement for him ever since.
This article is largely based on the work done
in collaboration with Tae-Sun Park,
to whom we owe our sincerest thanks.
The work of KK is partially supported by
the US National Science Foundation under grant number
PHY-0758114 and that of MR by the WCU Project
of the Korean Ministry of Education, Science and Technology
(R33-2008-000-10087-0).

\end{document}